# Energy self-sufficient systems for monitoring sewer networks

Simon Mathis[1], Juan-Mario Gruber[1], Christian Ebi[2], Simon Bloem[2], Jörg Rieckermann[2]
Frank Blumensaat[2, 3]

[1]Institute of Embedded Systems, ZHAW, Winterthur, Switzerland
[2]Urban Water Management Department (SWW), EAWAG, Dübendorf, Switzerland[2]
[3]Landesdirektion Sachsen, Dresden, Germany

Corresponding author: Juan-Mario Gruber, gruj@zhaw.ch

## Abstract

Underground infrastructure networks form the backbone of vital supply and disposal systems. However, they are under-monitored in comparison to their value. This is due, in large part, to the lack of energy supply for monitoring and data transmission. In this paper, we investigate a novel, energy harvesting system used to power underground sewer infrastructure monitoring networks. The system collects the required energy from ambient sources, such as temperature differences or residual light in sewer networks. A prototype was developed that could use either a thermoelectric generator (TEG) or a solar cell to capture energy needed to acquire and transmit ultrasonic water level data via LoRaWAN®. Real-world field trials were satisfactory and showed the potential power output, as well as, possibilities to improve the system. Using an extrapolation model, we proved that the developed solution could work reliably throughout the year.

## 1  Introduction

Underground infrastructure systems for water, electricity and other services form the backbone of our society. For example, underground drinking water and wastewater networks have provided reliable sanitation for the past 150 years. This advancement is considered to be an even greater medical achievement than the invention of antibiotics and vaccination [1]. In a similar fashion, district heating is a promising technology to mitigate the climate crisis [2]. Despite the value that these underground infrastructure systems bring to society, they are typically not monitored. This lack of data on system functioning is a problem because a malfunction leads not only to a decrease in function, but also to loss of economic value, etc. Monitoring could reduce this [3]. Although "Smart City"-type applications promise efficiency gains for underground infrastructure, the required technology for corresponding IoT (internet of things) sensor networks in the underground are largely lacking [4], [5]. A serious bottleneck is that, in the future, the operation of thousands of IoT-sensor nodes will require a huge number of batteries. This is unfortunate because batteries require (i) large amounts of energy and resources for production, (ii) disposal at the end of life and (iii) considerable manpower and operational resources for replacement in confined and hazardous underground spaces, such as sewers. Thus, to facilitate scalable IoT-applications for underground infrastructure, there is a need to locally generate energy and, ideally, create specifically-tailored self-sufficient monitoring systems. While indoor solar cells are efficient at collecting residual light, recent advances in thermal energy harvesting methods show that temperature differences of just 1.5-3 degrees Kelvin are sufficient to power sensors or actuators [6], [7]. Unfortunately, this technology has not yet been widely applied to underground infrastructure applications, such as sewer networks.

In this study, we present a self-sufficient, energy monitoring system for sewer networks. Presented as follows, our work has three main innovations. First, we present theoretical and practical considerations to design energy harvesting systems in underground infrastructure based on thermal differences and residual ambient light in the underground. Our design integrates all components of sensing, data transmission and energy supply, as well as fit-for-purpose considerations. Second, we design two prototypes for real-world tests. Finally, we demonstrate as a proof-of-concept that continuous monitoring and data transmission from sensor nodes in underground infrastructure is possible. Our results with a thermal-electric generator (TEG) and an underground photovoltaic solar cell (PV), located in a sewer manhole and a combined sewer overflow outlet channel, suggest that underground sensor nodes can operate on locally harvested power for several decades. The remainder of the paper is structured as follows. First, we present detailed background information on energy sources in underground infrastructure and argue why thermal energy and underground ambient light are the most promising sources. Second, we describe our experimental investigations in a sewer manhole and a combined sewer overflow outlet channel. Third, we describe how we use the experimental results to design our prototype, which harvests energy with a TEG or a PV and performs continuous monitoring with one (TEG) or two sensors (PV). Finally, we discuss our results as well as the optimization potential and draw our final conclusions.

## 2 Energy harvesting in underground water infrastructure

Underground water infrastructure is a major challenge for energy harvesting. In sewer systems, for example, the special circumstances limit the choice of energy sources that can be harvested reliably to operate a wireless sensor network. Floating objects can damage mechanical harvesters in the wastewater. Magnetic fields and radio frequency waves are highly dampened in the underground. Air flow is very weak in many sewer sections. In addition, the following practical design requirements were considered: i) the harvested power should be sufficient to drive ultrasound or pressure sensors for liquid level monitoring, the current workhorse in sewer surveillance, ii) data should be transmitted by an available low power wide area network , iii) the system should be simple to install (e.g., without long cables), and readily integrated into existing manholes and covers. Consequently, manhole covers with integrated solar cells and thermoelectric harvesting methods such as those used by [8] were ruled out because of the risk of damage when the manhole covers are lifted improperly.

In the following subsections, we describe i) a thermoelectric system that exploits the temperature difference between the wastewater and headspace temperatures, ii) a custom photovoltaic system that harvests the residual ambient light inside the manhole, and iii) an energy harvesting model, which can extrapolate the results from field measurements. The latter is very useful to predict the annual collectable energy for different hardware components and to select the most promising configuration.

### 2.1 Thermal energy

Domestic, commercial, and industrial wastewater can have a temperature difference of several degrees compared to the sewer headspace or wall (Figure 1). Heat recovery from wastewater is increasing in popularity [9], [10]; however, the potential varies depending on altitude, time of day, time of year, and catchment area of the drainage system [11]. Temperature measurements in the sewer over a period of 12 months in Eawag's urban water observatory (UWO) [12] show that seasonal variations affect both water and air temperatures similarly (Figure 1). The UWO is an experimental field laboratory of Eawag for testing new technologies in the fields of sensors and data transmission. These technologies will be used in the future for monitoring the condition of the sewage system. Battery-powered sensor nodes are already being used at the UWO, which transmit their data to an existing LoRaWAN® [13] network.

The existing thermal energy can be converted into electrical energy using thermoelectric generators (TEG). TEGs use the Seebeck effect to generate an electrical voltage from temperature differences [14]. Their biggest advantages are that they do not have mechanical components and are resistant to dirt. Since the output voltage of a TEG is in the millivolt range, a boost converter is used to adapt the output power. In order for a TEG to be able to transmit its maximum power to the booster, there must be power matching. For the extrapolation model, efficiency measurements were performed as a function of the applied input voltage. The power transferred to the booster was calculated in each case using the ratio of the internal resistance of the TEG and the impedance of the booster (see section 3.1).

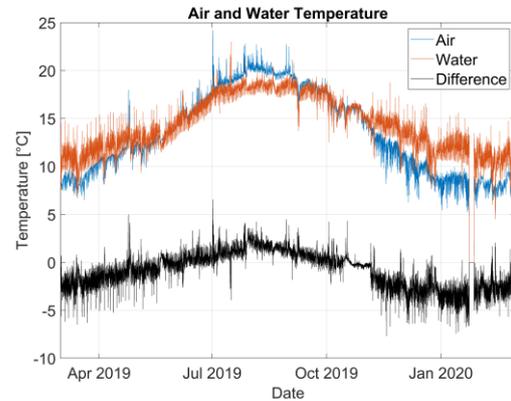

**Figure 1** Temperature measurements in the sewer in Eawag's urban water observatory show that seasonal variations affect both water and air temperatures similarly

### 2.2 Light energy

Although photovoltaics can be used wherever light enters the underground infrastructure, this energy source is not always present, or in very small amounts. Light only enters through small ventilation openings in the manhole cover or through outlet pipes into receiving waters. Similar to the TEG, the amount of harvestable light varies significantly throughout the day and the year. Nevertheless, our field trials show that the high efficiency of the solar cells can compensate for the low illuminance in the sewer.

With the solar cell setup, two reference measurements were made at different locations. By measuring the output power of different solar cell types under the same irradiation conditions, different solar cells can be directly considered in the model. The efficiencies of the converters were determined by varying the input voltage and input power. To predict the annual harvestable energy, we used local observations of the irradiation power (Figure 2).

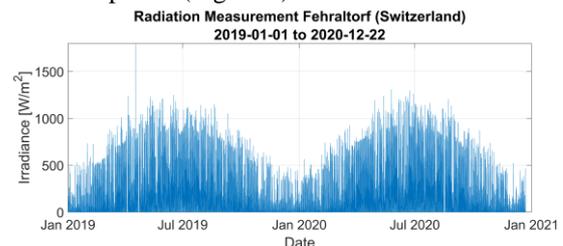

**Figure 2** Surface irradiation measurements in the test catchment Fehraltorf, CH show strong seasonal and daily variations superimposed by weather effects.

## 2.3 Extrapolation of energy harvesting potential

To assess the feasibility of our self-sufficient energy systems, we consider the power output over a whole year. We developed a model to extrapolate the reference measurements, collected over several days, to the period of a year (Figure 3) for both the TEG and the solar cell. In addition, the model also includes different harvesters and electronics such as TEG, solar cells and DC-DC converters. Thus, the model also simulates different hardware combinations. Despite its simplicity, the model also considers the losses of power conversion and delivers predictions for the annual collectable energy.

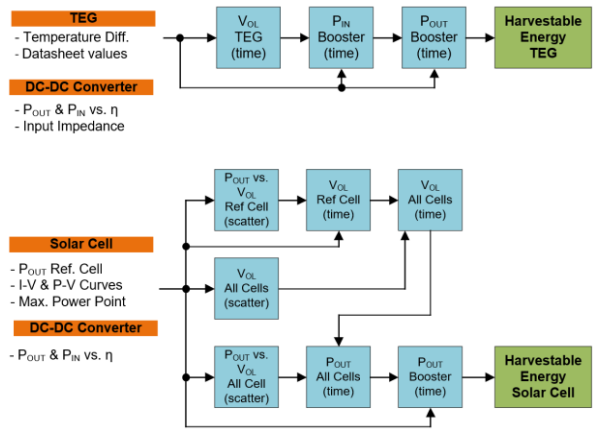

**Figure 3** Extrapolation model to determine the annual energy that can be collected from the TEG and solar cell.

## 3 Experimental investigations in a real-world sewer network

Various function and performance tests were carried out with the two setups for both the TEG and the solar cell. The Urban Water Observatory (UWO) in Fehraltorf [15] was selected as the test environment.

### 3.1 Harvesting thermal differences

The thermal energy harvesting potential is analyzed using a reference TEG (TEG2-40-40-4.7/100). A mechanical structure thermally couples the TEG to the wastewater and to the sewer wall. The mechanical structure consists of a 3-D printed part, aluminum plate and a traditional spreader ring for the installation of submerged sensors. The top two images in Figure 8 show the construction of the 3D part with the aluminum profiles. The TEG is in the middle of the 3D print and is pressed onto the two aluminum profiles via two PVC plates with four screws. The screwed assembly is attached to the spreader ring with adhesive tape during assembly. The finished construction can be seen in the graphic below right in Figure 8 to improve the heat transfer to the channel wall, one side of the aluminum profile has been fitted with thermal pads. The clamping ring is then mounted inside the canal on the canal wall and pressed against the sewer wall using a spreader mechanism (bottom left image in Figure 4).

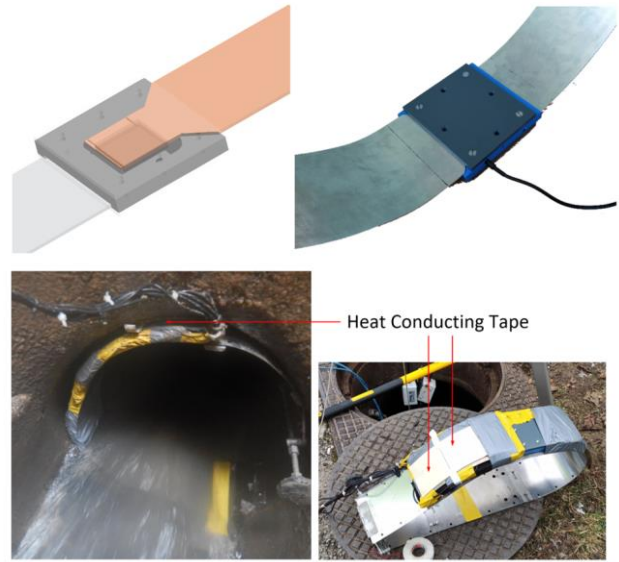

**Figure 4** Setup with a TEG for measuring the available power in a sewer pipe

To investigate the potential for energy harvesting available power for analysis purposes, the transmitter unit measures and transmits the TEG voltage across a reference resistor under load and at no load. Because the load and internal resistance of the TEG are approximately equal, the power at the load resistance can be calculated directly from the measured voltage. Four additional temperature sensors were installed i) at two different positions of the wall, ii) in the wastewater flow and iii) in the headspace. In Figure 5 48h of temperature dynamics is shown as an example.

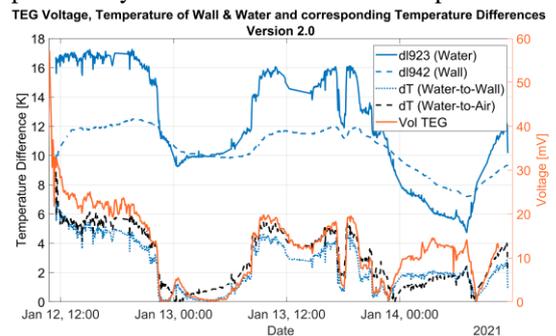

**Figure 5** Temperatures and temperature differences between channel wall and wastewater. The open-circuit voltage of the TEG shows a strong relation to the temperature difference

It can be seen that the output voltage of the TEG follows the course of the temperature differences of wastewater-channel wall and wastewater-air. Although a maximum difference of about 8 K was measured between the media, the TEG could reach a maximum voltage of about 30 mV.
Using the extrapolation model (**Figure 3**), the expected collectable energy for the reference TEG was calculated using the measurements at the TEG and three boosters LTC3108, Mercury and EM8900. Figure 6 shows the results of these calculations. The curves indicate the expected annual collectable energy as a function of the efficiency of thermal coupling between TEG and media.

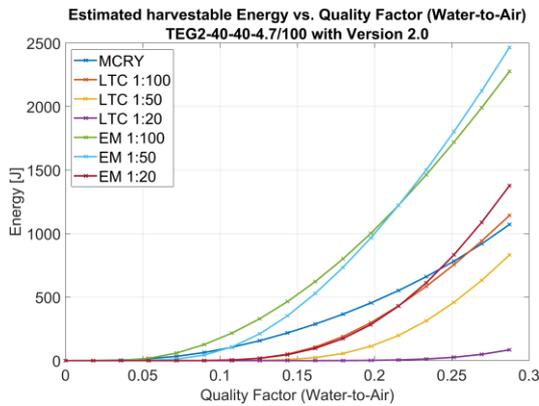

**Figure 6** Expected collectable energy with the reference TEG in combination with the three boosters LTC3108, Mercury and EM8900 as a function of the efficiency of thermal coupling

The quality factor (Figure 6) is the ratio between the temperature difference at the TEG and the temperature difference of the thermally connected media. For example, if the media have a difference of 5 K, but only 0.5 K drop at the TEG, the setup has an efficiency or quality factor of thermal coupling of 0.1. In combination with the EM8900, it is expected that the reference TEG can collect the most energy. Among other things, this calculation becomes more understandable because this booster has a more optimal input impedance for the TEG compared to the other converters and already operates at 5 mV. With the spreader ring prototype a factor of 0.073 and a collectable energy of about 60 J was achieved.

### 3.2 Harvesting residual ambient light

For the use of residual light with a solar cell, a measuring setup was installed in an outlet channel (Figure 7) as well as directly below the ventilation hole of a manhole cover (Figure 8). The measurement setup consists of two battery-powered subsystems, a measurement unit and a transmission unit. In addition, a lux meter type HD450 was installed parallel to the solar cell in the outlet tunnel for 24 hours.

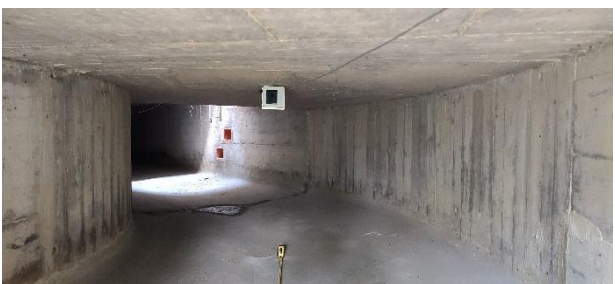

**Figure 7** Setup for measuring the available power of a solar cell in an combined sewer overflow outlet pipe (viewed from watercourse exit)

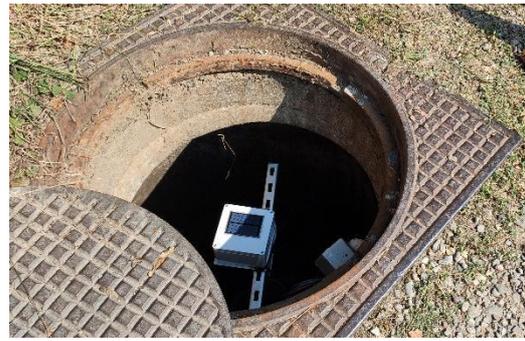

**Figure 8** Prototype mounted directly below the ventilation hole of a manhole cover

The measuring unit of the setup includes a reference solar cell and a specially developed measuring circuit for determining the MPP of the solar cell during operation. The transmitter unit is a commercial LoRaWAN® node (DL-DLR2-005, Decentlab, Switzerland) that provides a 24 bit analog-to-digital converter and periodically transmits the measured voltage. Both subsystems are independently supplied with energy via batteries. The microcontroller in the measuring circuit measures the power every 30 seconds and forwards it to the transmitter unit in the form of a voltage. The transmitter unit measures this voltage every two minutes and then sends it via a gateway to a central server for further processing. The graphs in Figure 9 show the measured maximum output of the solar cell in mW and the irradiation in $W/m^2$ from a weather station (WS700, Lufft, Germany) about 500 m away.

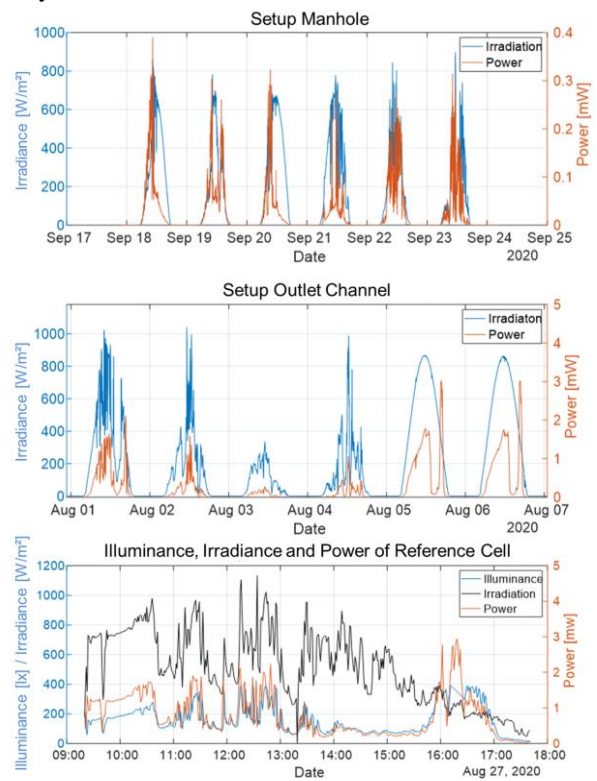

**Figure 9** Measured performance of the reference solar cells and measured irradiance of the WS700 weather station at the nearby wastewater treatment plant in Fehraltorf, Switzerland

The measurement at the outlet channel is a six-day excerpt from a total of 34 days. The measurement setup in the descent manhole was only in operation for six days. When measuring in the outlet duct, cloudy (1st to 4th August inclusive) and sunny days (5th and 6th August) can be seen in the irradiation. On a sunny day, the irradiation has a steady course and forms the shape of a Gaussian bell curve. On rainy days, the influence of cloud cover is noticeable both in the intensity and in the discontinuous form of the insolation. The non-steady course of the power is related to the indirect irradiation or the angle of irradiation in the duct. The dip in the second half of the day is caused by an obstacle that dampens the irradiation into the manhole. In general, the solar cell generates a peak power around 0.3 mW. Here, too, sunnier and cloudy days can be easily recognized by the irradiation. The 24-hour measurement with the HD450 shows that the solar cell$_{RMS}$ generates a power of approx. 0.78 mW$_{RMS}$.

Based on the measurements on the solar cell in Figure 9 and voltage converters BQ25570 and SPV1050, the expected annual energy of the four different solar cells was predicted using the extrapolation model (Section 2.3) . The results of these calculations are shown in the diagram in Figure 11 for the manhole installation. Installed in the outlet channel, the calculated energy is about 3-4 times higher on average. Both models were calculated for a maximum voltage of 5.3 V at the storage. In this mode of operation, in the manhole (Figure 8), both converters SPV1050 and BQ25570 are expected to be able to collect more than 8 kJ with the most powerful solar cell (SM531). In this scenario, the BQ25570 harvests almost 20% more energy due to the higher efficiency over the entire input voltage range.

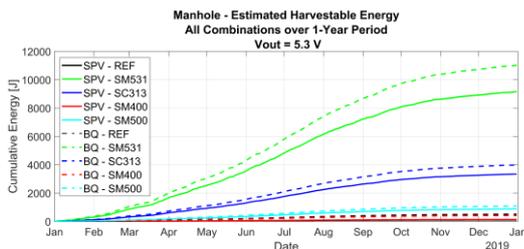

**Figure 10** Extrapolated course of the collectable energy of the four evaluated solar cells and two voltage converters

## 4 Prototype for real world tests

For real world investigations, a prototype has been developed to harvest thermal and solar energy from sewers and perform continuous monitoring with one (TEG) or two (PV) sensors (Figure 12). It consists of a PCB with the electrical components and an IP66 housing, where a solar cell can be mounted onto the cover (as shown in Figure 7 or Figure 8).

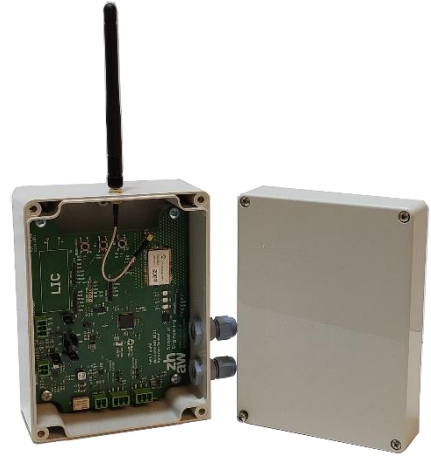

**Figure 11** Prototype for monitoring infrastructure in urban water management includes energy harvesting, data acquisition, data analysis and data transfer in a waterproof housing. A solar cell was mounted to the cover.

### 4.1 System concept

To observe the spatio-temporal behavior of sewers, wireless sensor networks are ideally suited. A single node harvests ambient energy, reads out measurement data from the sensors and transmits measurement data to the network. Figure 12 shows the block diagram of such a node. The electronics are divided into the energy harvesting system and the measurement and data processing system. The harvester converts the energy from the environment into electrical energy and emits an output voltage. Feeding the system directly from this voltage is not always possible due to varying environmental conditions. A voltage converter transforms the voltage level of the harvester to a level suitable for the energy storage device. This level must be below the maximum allowable voltage of the energy storage device. A power control circuit prevents overcharging and deep discharging of the storage device.

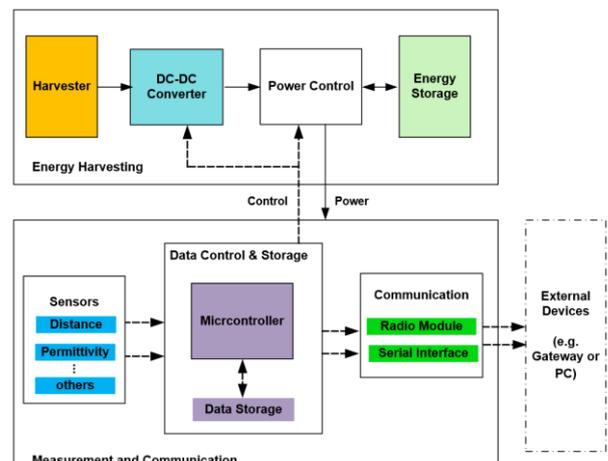

**Figure 12** Block diagram and subsystems of the self-sufficient sensor node

Our prototype monitors water levels via the sensor (see section 3.3.) and transmits the acquired measurement data to the gateway via the wireless communication interface. Transmitting data from underground requires a high link budget due to signal attenuation [4]. To reduce the increased risk of data loss, all measurement data is stored on a non-volatile on-board data memory.

### 4.2 Energy harvesting

Our prototype can harvest energy either with a TEG or a solar cell. The selected thermoelectric generator type (TEG2-40-40-4.7/100, Eureka, Germany) has a thermal force of 53 mV/K and a resistance of 1.5 Ω. It is combined with a boost converter (EM8900, EM Microelectronic, Switzerland). This booster operates from an input voltage of 5 mV. In combination with a $C_{fb}$ of 100 pF and a 1:100 transformer, it has an input impedance of 2 Ω at 40 μW input and 3 V output and an efficiency of 50 %. This corresponds to a power transfer factor of 97.9 % in combination with the selected TEG. The resulting power at the output of the booster is 19.58 μW.

For the ambient light powered solution, a solar cell (313070004, Seeed Studio Ltd.) was selected with an area of 3850 $mm^2$. As a voltage converter, a SPV1050 buck-boost converter is used because with 18 V input voltage, it offers sufficient reserve for solar cells with a higher MPP voltage.

The collected energy is stored in a lithium-ion capacitor [16]. This type of hybrid capacitor combines the advantages of a capacitor (number of cycles) with the advantages of a lithium battery (higher energy density and lower self-discharge). Since an LIC is destroyed by deep discharge below 2.2 V, a reset circuit is used. This circuit disconnects the LIC from the sensor-microcontroller part via a load switch when the voltage drops below 2.25 V. As soon as the voltage exceeds the level of 2.45V, the LIC is reconnected. The total current consumption of this circuit is between 183 nA and 226 nA, depending on the output state of the comparator.

### 4.3 Sensor technology and data processing

Ultrasonic distance sensors or membrane pressure sensors are frequently used for level measurements. Float switches or capacitive conductivity sensors can be used to detect water at overflow weirs. All of these sensors have a low operating voltage of around 3V or lower. Energy consumption is in the range of a few hundred microjoules to approximately 10 mJ [17] per measurement. Outside of these measurement intervals, the sensor is completely disconnected from the power supply and requires no power.

A microcontroller controls data acquisition, processing, storage and transmission. Modern microcontrollers have operating voltages in the range of 1.8 V and standby current consumption in the three-digit nano-ampere [18] range. The microcontroller coordinates the operations on the node and is also part of the energy management. An STM32L073 was selected here, because it has various low-power modes and achieves a current consumption of 0.86 μA in stop mode. The internal real-time clock (RTC) can wake up the microcontroller without an external signal. The measurement data is transmitted via LoRaWAN® with the module type RN2483. For longer deep sleep phases, the LoRa module is disconnected from the power supply via a TPS22860 load switch. This switch has a leakage current of about 12nA [19].

### 4.4 Energy balance

Due to the limited energy available, the operation of the sensor system is energy-optimized. For this purpose it is split into two phases with different power consumption. In the inactive phase, the microcontroller and all other subsystems are in deep sleep or switched off with minimal power consumption. In the active phase, the system performs the primary tasks such as measuring, reading memory or communication with the gateway and data transmission. In this phase, the required energy depends on time duration and the power consumption of the tasks. In order for self-sufficient operation to be possible, it must be possible to collect enough energy to compensate for the required amount of energy for the inactive phase. Any excess energy is then available for the active phase.

The extrapolation models were used to calculate the annual amount of harvested energy within one year. According to the model, the selected TEG2-40-40-4.7/100, combined with the EM8900 and the realized setup, is expected to provide an energy of 60 J per year. The lowest power solar cell (SM400, 495 mm2) in combination with the SPV1050 at 5.3 V charge voltage achieves a projected energy of 1 kJ per year according to the model. The larger solar cells provide up to 36 kJ per year in our manhole, depending on whether they are installed in the outlet duct or descent manhole.

For the energy balance of the system, the power consumption of the developed prototype is divided into inactive and active phases. Both configurations require the same energy in the active phase. An active cycle, which consists of waking up, measuring, reading and writing FRAM and sending the data via LoRa, requires about 271 mJ. For the inactive phase, a differentiation was made between the system with TEG and solar cell, since the converters require a different standby current. Using a Power Analyzer N6705B, a standby power consumption of 4.5 μW for the TEG-system and 7 μW for the solar system has been measured (complete system with quiescent current of the converter and microcontroller in standby mode). This results in approximately 141 J and 220 J energy needed per year, which is too much basic consumption for the TEG.

To power the system with the TEG, it is necessary to modify the power management. The average power consumption can be reduced, by lowering the switch-on threshold of the reset circuit to 2.3 V. Using the smallest available LIC with 10F, the sensor-microcontroller system is permanently disconnected from the battery until enough energy is available to switch to the active phase. During this inactive phase, where the sensor-microcontroller system is disconnected, the average power loss is reduced to about 1.15 µW and annual 36.16 J respectively. Thus, the system has 23.83 J per year or 0.45 J per week available for the active phase. Measuring and transmitting requires about 0.27 J. It is therefore possible to send a measurement data package weekly, given that the TEG has a weekly energy buffer of 0.18 J.

With all solar cells except the SM400K10L, the minimum required energy $W_{SLEEP}$ can be compensated. In the active phase, between 2446 and 118261 measurement and transmission cycles can be performed annually, depending on the type. This corresponds to a temporal resolution of 215 min and 5 min, which is sufficient for most dry weather monitoring tasks. For rainy weather, flow processes are much more dynamic and a monitoring interval of 30s or 1 min is desirable.

## 5  Optimization potential and discussion

Our prototypes of energy self-sufficient LoRaWAN® nodes for sewer monitoring are very promising, because continuous water level monitoring is possible. As we used a small solar cell and a simple TEG, there is still significant room for improvement. In the following section, we discuss i) the energy harvesting design, ii) further integration of the individual components of the consumer side, iii) potential energy savings as well as, iv) practical considerations.

For energy harvesting, an optimized design of our clamping ring can improve the thermal coupling to the wall and the wastewater. By minimizing the thermal resistance, the heat flow through the TEG can be significantly increased. For the PV, a slightly higher efficiency could be achieved by using the Texas Instruments BQ25570 as a boost converter instead of the SPV 1050. For this, the solar module voltage must be limited to less than 5 V. To avoid the build-up of dirt directly on the PV cell, an installation perpendicular to the manhole cover, possibly using a set of mirrors which direct ambient light towards the cell, could be promising.

On the consumer side, first, an external real-time clock (RTC) can be used to reduce static power consumption. In the inactive phase, only the RTC is connected with a current consumption in the nano-ampere range. It could then periodically enable the sensor-microcontroller part during the active phase. Second, microcontroller or memory components with increased power-saving could be used. Third, one could reduce the power required for communication. For LoRaWAN®, the energy required for a transmission of 2 to 4 bytes of user data is in the range of 40 mJ and for NB-IoT, it is around 64 mJ [20]. However, under optimal conditions, energy consumption per transmitted data packet has been reduced to 1.2 mJ [21] and 1.9 mJ [22] by tuning transmission parameters.

As rainfall-runoff in sewers is more dynamic than dry weather flows, an "adaptive monitoring" mode could save further energy. In this mode, the active phase would not be triggered by an RTC, but by an external signal, e.g. from an internet weather service, weather radar or upstream node in the sewer network. During dry weather (ca. 90% of the time), a monitoring interval of 30 min is likely sufficient, instead of 5 min intervals.

Finally, LoRaWAN® is not suitable for all aspects of "Smart City"-type applications. For example, in Real Time Control applications, actuators increase or limit individual flows in certain locations to improve the performance across the entire sewer network. Here, the reliability of packet delivery when transmitting from range-critical locations is probably not sufficient [5] and other modes of data transmission are needed.

## 6  Conclusions

In this study, it was shown that a sensor can be operated in underground infrastructure using energy harvesting without an external power supply or batteries. For this purpose, residual light and temperature differences were considered as possible energy sources. The methods tested in practice contribute to the enabling of virtually maintenance-free monitoring of underground and difficult-to-access structures.

Using prototypes that have yet to be optimized, it was shown that the residual light in the outlet channel or the thermal differences are sufficient to operate a wireless sensor throughout the year. The energy balance shows that a temporal resolution of up to 5 minutes can be achieved when using suitable solar cells. When thermal differences are used as an energy source and data is sent weekly, there is reserve for additional alarm messages.

The study shows further potential for optimization, especially with respect to thermal energy conversion. Future work will focus on reducing the standby consumption of the system and optimizing the harvesting performance by increasing the temperature difference at the TEG.

## 7  Acknowledgement

The authors would like to thank Lauren Cook from the Urban Water Management Department at Eawag for proofreading the English version and for her helpful comments.